\documentclass{aa}
\usepackage{txfonts}
\usepackage{natbib}
\bibpunct{(}{)}{;}{a}{}{,}
\def\SB9{$S\!_{B^9}$}
\usepackage[dvips]{graphicx}
\newcommand\ignore[1]{} 
\begin{document}
\title{Candidate spectroscopic binaries in the Sloan Digital Sky Survey}

\author{D.~Pourbaix\inst{1,2}\fnmsep\thanks{Research Associate, F.N.R.S.,
Belgium}, G.~R.~Knapp\inst{1}, P.~Szkody\inst{3}, {\v Z}.~Ivezi\'c\inst{3},
S.~J.~Kleinman\inst{4},
D.~Long\inst{4}, S.~A.~Snedden\inst{4}, A.~Nitta\inst{4},
M.~Harvanek\inst{4}, J~ Krzesinski\inst{4,5}, 
H.~J.~Brewington\inst{4}, J.~C.~Barentine\inst{4}, 
E.~H.~Neilsen\inst{6}, \& J.~Brinkmann\inst{4}}

\offprints{D. Pourbaix (ULB)}

\institute{Department of Astrophysical Sciences, Princeton University,
Princeton, NJ 08544-1001, USA; pourbaix, gk@astro.princeton.edu
\and 
Institut d'Astronomie et d'Astrophysique, Universit\'e
Libre de Bruxelles, CP. 226, Boulevard du Triomphe, B-1050
Bruxelles, Belgium; pourbaix@astro.ulb.ac.be
\and
Department of Astronomy, University of Washington, Box 351580,
Seattle, WA 98195; szkody, ivezic@astro.washington.edu
\and
Apache Point Observatory, P.O. Box 59, Sunspot, NM 88349; sjnk,
long, snedden, ank, harvanek, jurek, hbrewington, jcb, jb@apo.nmsu.edu
\and
Mt. Suhora Observatory, Cracow Pedagogical University, ul. Podchorazych 2,
30-084 Cracow, Poland
\and
Fermi National Accelerator Laboratory, P.O. Box 500, Batavia,
IL 60510, neilsen@fnal.gov
}
\date{Received date; accepted 21/08/2005} 
 
\authorrunning{Pourbaix et al.}
\titlerunning{Spectroscopic binaries in SDSS}

\abstract{
We have examined the radial velocity data for stars spectroscopically
observed by the
Sloan Digital Sky Survey (SDSS) more than once to investigate the 
incidence of spectroscopic binaries, and to evaluate the accuracy of the
SDSS stellar radial velocities.  We find agreement between the
fraction of stars with significant velocity variations and the 
expected fraction of binary stars in the halo and thick disk
populations.  The observations produce a list of 675 possible
new spectroscopic binary stars and orbits for eight of them.
}
\maketitle

\section{Introduction}

This paper is based on an investigation of the radial velocity accuracy
obtainable for stars observed in the Sloan Digital Sky Survey (SDSS)
which we carried out in support of current and upcoming observations
of Galactic stellar kinematics using SDSS spectroscopy \citep{Beers-2004:a}.
In the course of this investigation, we were able to identify a large
number of confirmed and candidate binary stars. While SDSS data have been used
in several recent studies of binary stars,
in particular of dwarf M-white dwarf pairs
and cataclysmic variables \citep{Szkody-2002:a,Szkody-2003:a,Szkody-2003:b,
Szkody-2004:a,Szkody-2005:a,Raymond-2003:a,Pourbaix-2004:a,Smolcic-2004:a},
this represents the first detection of spectroscopic binaries in SDSS. 
This paper presents both of these results.
The SDSS data
are described in Sect. \ref{Sect:SDSSspec} and \ref{Sect:RVerr}, 
and in Sect.~\ref{Sect:accuracy} we compare
velocities observed at different epochs and examine their precision.
Almost all of the stars with repeat observations
have been observed only twice, and we analyze the distribution of velocity 
differences to derive the binary fraction. In a small number of cases (19)
there are sufficient observations that a spectroscopic orbit can be derived, 
but only eight of these orbits prove to be robust.  These objects are discussed
in Sect. \ref{Sect:orbits}, where the full list of possible binary stars is 
also described.

\section{The Sloan Digital Sky Survey}\label{Sect:SDSSspec}

The Sloan Digital Sky Survey (SDSS) is a 5-band photometric survey of
about 10\,000 square degrees of the Northern sky to a depth of about
22.5 ($r$ magnitude, point source) and a concurrent
redshift survey of up to a million galaxies and 100\,000 quasars
selected from the imaging survey \citep{York-2000:a}.  The primary 
science goals of the project are to provide the data to investigate the
large scale structure of the Universe and other extragalactic science.
The photometric data are acquired almost simultaneously in five bands,
$u$, $g$, $r$, $i$, and $z$, centered at approximate effective wavelengths
of 3551, 4686, 6166, 7480 and 8932 \AA\ \citep{Fukugita-1996:a} using
a large-format CCD camera \citep{Gunn-1998:a} mounted on a dedicated 
2.5 meter telescope at the Apache Point Observatory (APO) in New Mexico.
The imaging data are automatically reduced through a series of software
pipelines which find and measure objects and provide photometric and
astrometric calibrations to produce a catalogue of objects with
calibrated magnitudes, positions and structure information
\citep{Lupton-2001:a,Lupton-2003:a,Pier-2003:a,Ivezic-2004:a}.
The instrumental fluxes are calibrated via a network of primary and
secondary stellar flux standards to $\rm AB_{\nu}$ magnitudes 
\citep{Oke-1983:a,Fukugita-1996:a,Hogg-2001:a,Smith-2002:b} which 
are accurate to about 1\% in $g,r$, and $i$, 3\% in $u$ and 2\% in $z$
for bright ($<$ 20 mag) point sources.  The bright magnitude limit
is about 14.  Absolute positions are accurate to about 50 mas  in each 
coordinate \citep{Pier-2003:a}.

Targets for spectroscopy are selected from the imaging data on the basis 
of their photometric properties. As well as the primary SDSS targets,
stars in many different locations of color-magnitude space are selected
to serve as spectrophotometric standards and to provide backup
targets in regions of low galaxy density. The target objects are mapped
\citep{Blanton-2003:a} onto aluminum $3^{\circ}$ diameter fiber plug
plates which feed the spectrographs.  The pair of dual fiber-fed 
spectrographs \citep{Uomoto-1999:a} can observe 640 spectra at one 
time with a wavelength coverage of 3800 - 9200 \AA\ and a resolution 
of 1800 to 2100. The fibers subtend an aperture of 3\arcsec\ on the sky.
The spectroscopic observations usually consist of three fifteen-minute
exposures per spectroscopic plate.

The data are optimally extracted from the CCD images, flat-fielded
and wavelength calibrated using
arc spectra and the night-sky lines observed on
the plate.   A mean sky spectrum is subtracted
from each object spectrum, which is then flux-calibrated 
with respect to the F star calibration spectra. Regions of the spectrum
with bad data (for example, in the immediate wavelength vicinity of strong
night sky lines) are flagged so that they will not be used in subsequent
analysis. The
resulting calibrated 1D spectra are fit to a series of
templates of galaxies, quasars and stars to derive the spectral
classification, redshift and redshift error of each object
(D. Schlegel, in preparation). 

There are several extensive libraries of stellar spectra, which can be used
as templates for spectral type and radial velocity fitting.   Our work
began with the flux-calibrated spectra from the Elodie library \citep{Prugniel-2001:a,Moultaka-2004:a}. However, since the Elodie spectra do
not have as large wavelength coverage as do the SDSS spectra, they are
not ideal for direct use as templates. The stellar templates were therefore
used as follows. First, the Elodie spectra were matched against the SDSS
spectra and used to extract spectra with a good signal-to-noise ratio ($>$
15 per spectral resolution element), to assign best-fit spectral types and
to correct the SDSS spectrum to a velocity of 0 km/s (the Elodie spectral
library has systematic errors less than 1 km/s). Next, the calibrated
and typed SDSS spectra were used to select representative
spectra of all observed spectral types. These spectra were used 
to construct templates by combining individual spectra and defining
the principle components \citep[see][]{Heyer-1997:a} to produce a set of
templates simulating a wide range of effective temperature, gravity and 
metallicity. These templates were then fit to all SDSS spectra directly in
flux density-wavelength space, using $\chi^2$ minimization to assign
the most likely spectral type and redshift. For the subset of objects
classified as stars, the assigned spectral type of each star is that
of the best fit stellar template, and the radial velocity is calculated
from the redshift necessary to bring the template spectrum and object
spectrum into optimum alignment in wavelength space.

This process produces both a radial velocity and a radial velocity error
for the best fit template, but there is a second source of radial velocity
error, that arising from template mismatch. To evaluate this error, each
SDSS stellar spectrum is fit to all 900 Elodie spectra and the standard
deviation of the radial velocities of the 12 best-fit templates computed.
This quantity can sometimes be much larger than the random error if
there is significant template mismatch, and it is included in the
total radial velocity error.

For the particular application discussed here, the investigation of radial
velocity changes for a given star which are larger than the random errors
and therefore may be due to binary motion, the errors introduced by
template mismatch are less important, since the same template will be fitted
to the stellar spectrum for each epoch of observation (except in very
rare cases, such as that shown in Fig.~\ref{Fig:spectCV} below), and indeed 
this was checked for the multiply-observed stars analyzed in the next sections.
Some stars, however, may have no good template available because their
lines are broadened by rapid rotation, and the velocity errors will not
be well determined. Fortunately, such stars are very seldom found in
the SDSS data base.

The SDSS data are described in the data release papers 
by \citet{Abazajian-2003:a,Abazajian-2004:a,Abazajian-2005:a} 
and documented at the web sites listed therein and at 
{\bf http://www.sdss.org},
where the sky coverage of the SDSS observations is also described.

\section{Multiple radial velocity observations}\label{Sect:RVerr}

The objects which were both targeted and spectroscopically classified as stars
were extracted from the spectroscopic data base using all data obtained up to 
January 4, 2005. 
Because the SDSS observes the high-latitude sky and has a bright limit of
about 14, most of the stars observed lie in the Galactic halo and thick disk.
In particular, the F subdwarfs, sixteen of which are observed for every plate
to act as photometric standards,
lie in the halo.
A small fraction of stars has been observed spectroscopically
more than once, either to provide quality
checks for the data or occasionally by chance (this is especially true for the
spectrophotometric standard stars). The data for stars observed more than once,
10\,647 in all, were then identified. 

\begin{figure*}[htb]
\resizebox{0.49\hsize}{!}{\includegraphics{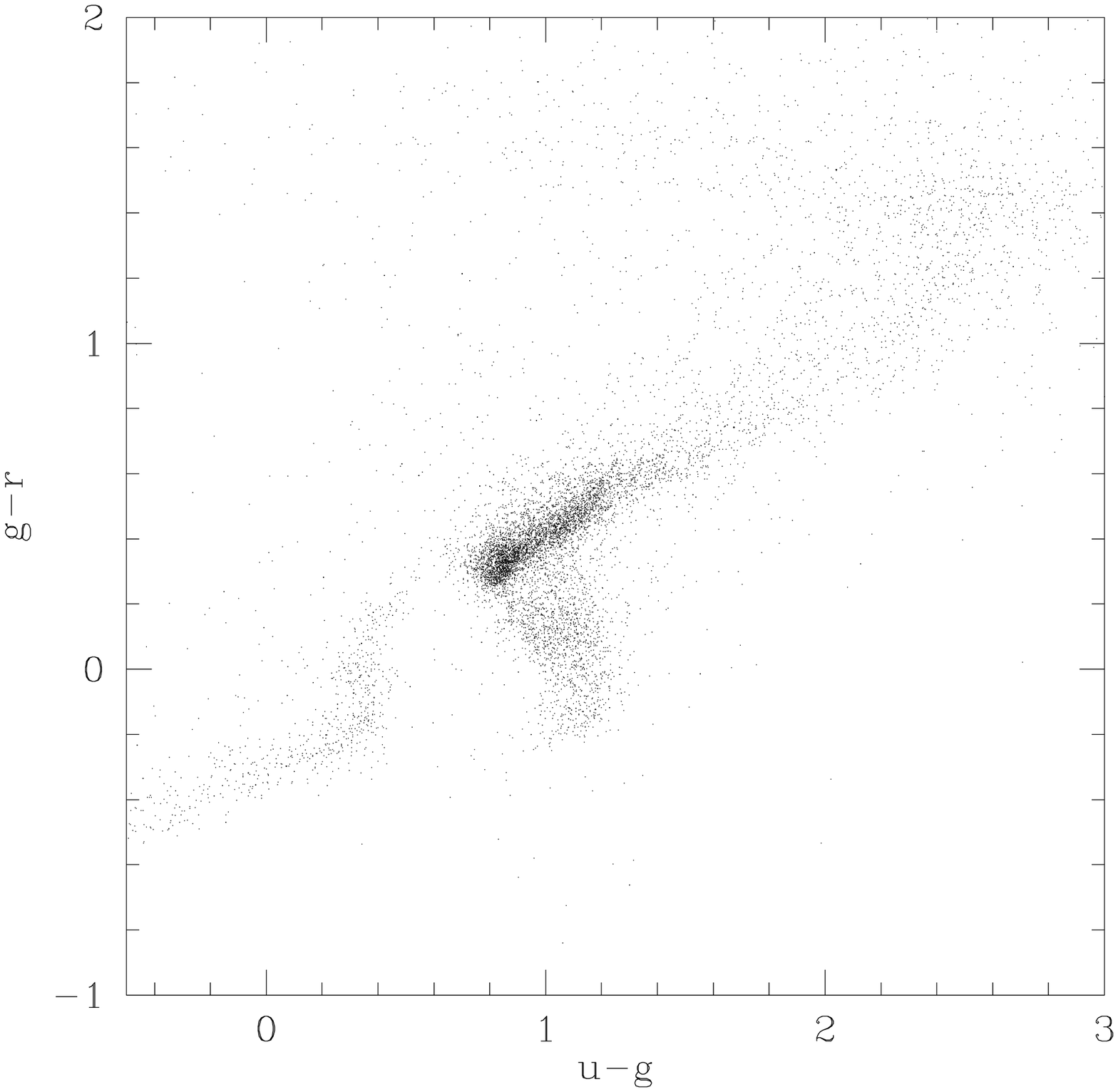}}
\resizebox{0.49\hsize}{!}{\includegraphics{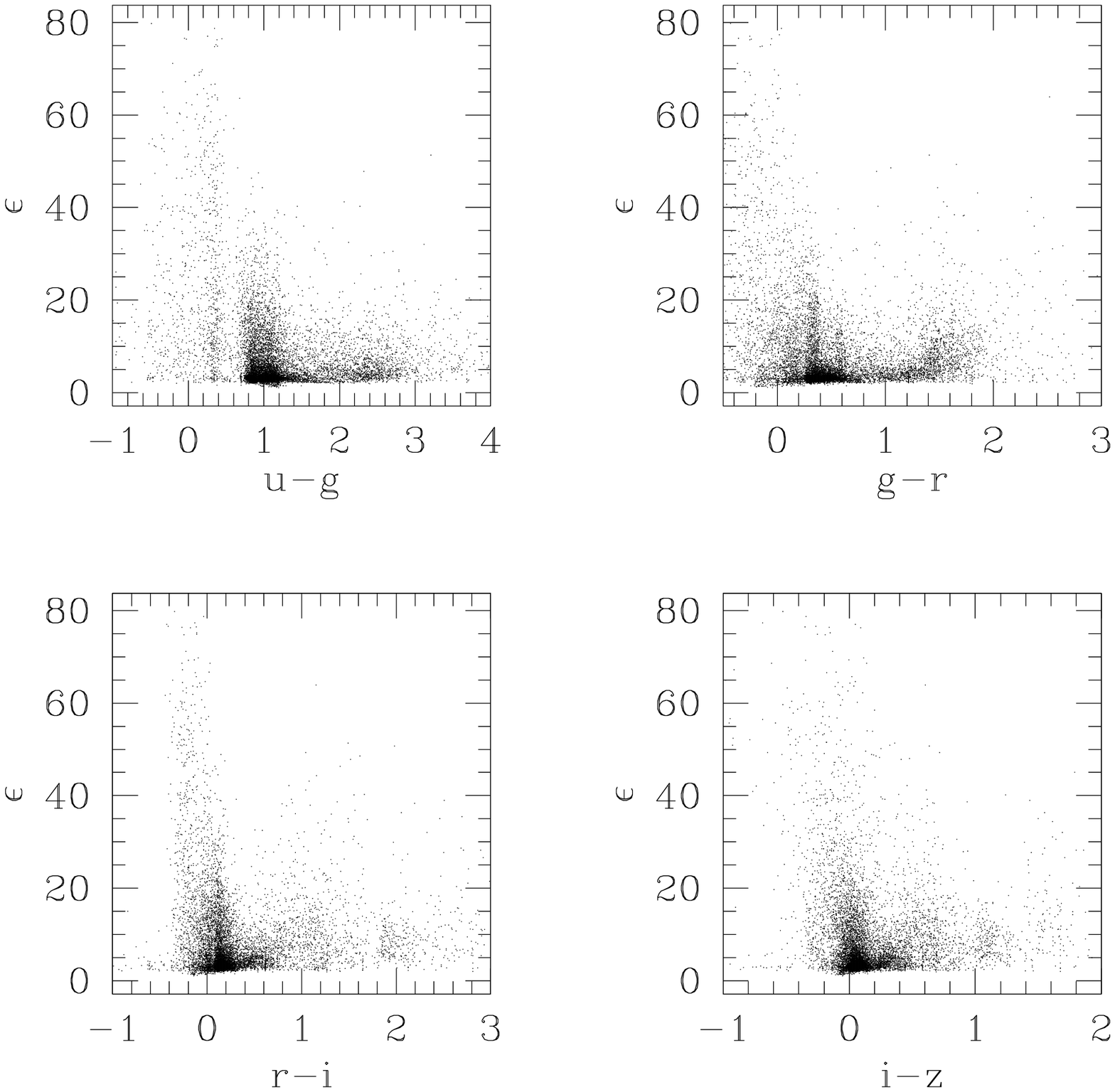}}
\caption[]{\label{Fig:CCandstddev2pts}(Left panel) Color-color diagram of the 10\,647 stars with multiple SDSS spectral observations. The colors are uncorrected for interstellar extinction.  (Right Panel) Mean radial velocity uncertainty (km/s) versus color.}
\end{figure*}

The $u-g$ vs $g-r$ color-color diagram for these stars
is shown in the left panel of Fig. ~\ref{Fig:CCandstddev2pts}. The stars cover
essentially the full range of stellar colors observed by SDSS 
\citep[see][]{Finlator-2000:a}.  The large number of stars in the F subdwarf 
region occurs because of the use of these objects as spectrophotometric 
standards;
they are the only stars for which SDSS observes a representative
sample.

For a given multiply-observed
star, we have a set of observations of $V \pm \epsilon$, the heliocentric
radial velocity and its uncertainty.  The right panel of Fig.~\ref{Fig:CCandstddev2pts} shows the mean $\epsilon$ versus color for the sample of stars. As expected, the uncertainty of the radial velocities is color-dependent, with large values for $u-g<0.5$. This occurs because many
of the blue stars are white dwarfs, whose very broad lines preclude the 
measurement of the radial velocity to accuracies of 10 $\rm km~s^{-1}$,
which are typical for the observations of main sequence stars (see below).
Note also that the 45-minute exposure time for the spectroscopic observations
will lead to broadening of the spectral lines of stars with rapidly-
varying velocities, and that additional uncertainties can be introduced by
the assumption of a single radial velocity for all spectral lines. 

\section{Velocity Accuracy and Incidence of Binary Stars}\label{Sect:accuracy}

\begin{figure}[htb]
\resizebox{0.49\hsize}{!}{\includegraphics{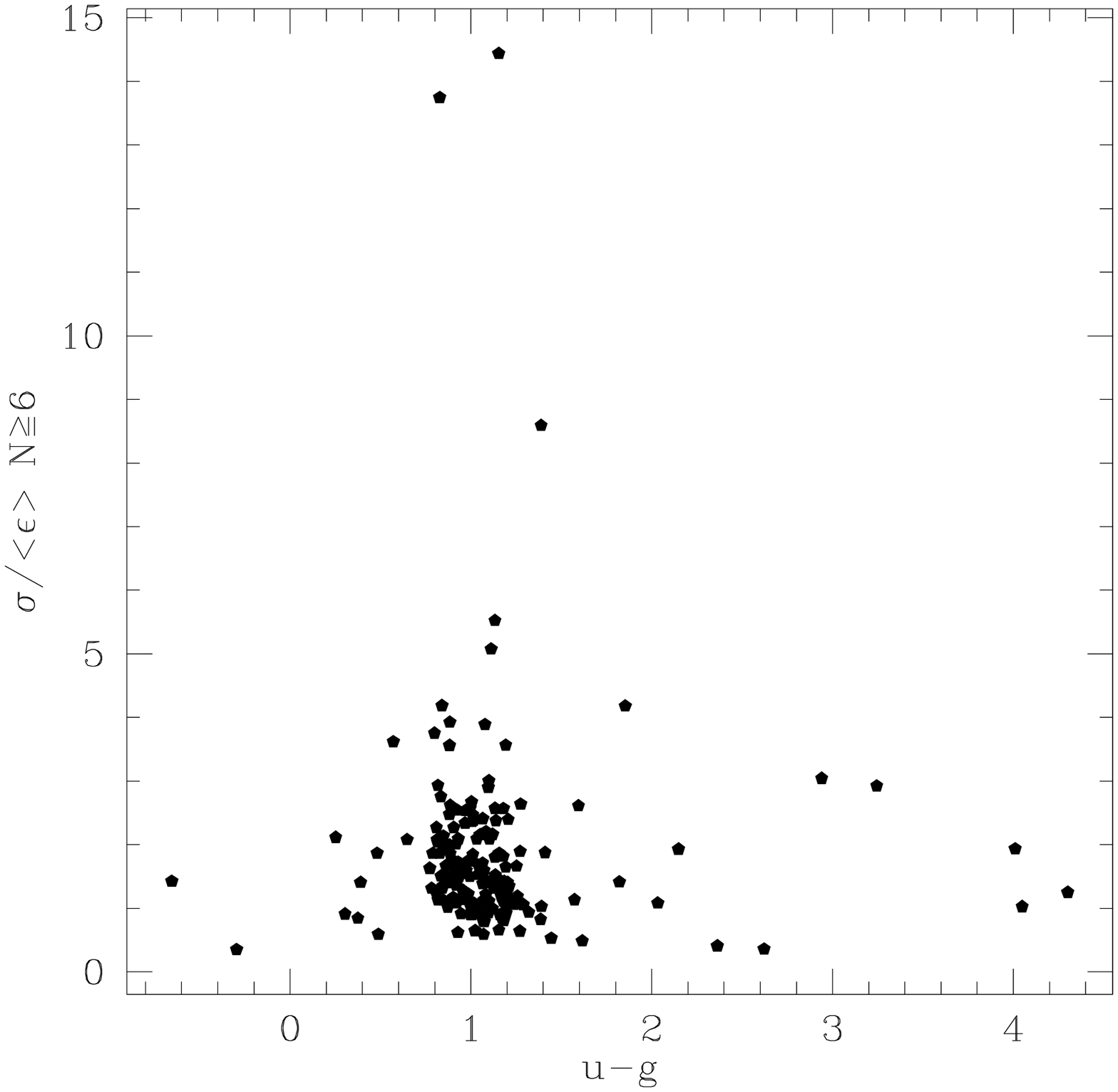}}
\resizebox{0.49\hsize}{!}{\includegraphics{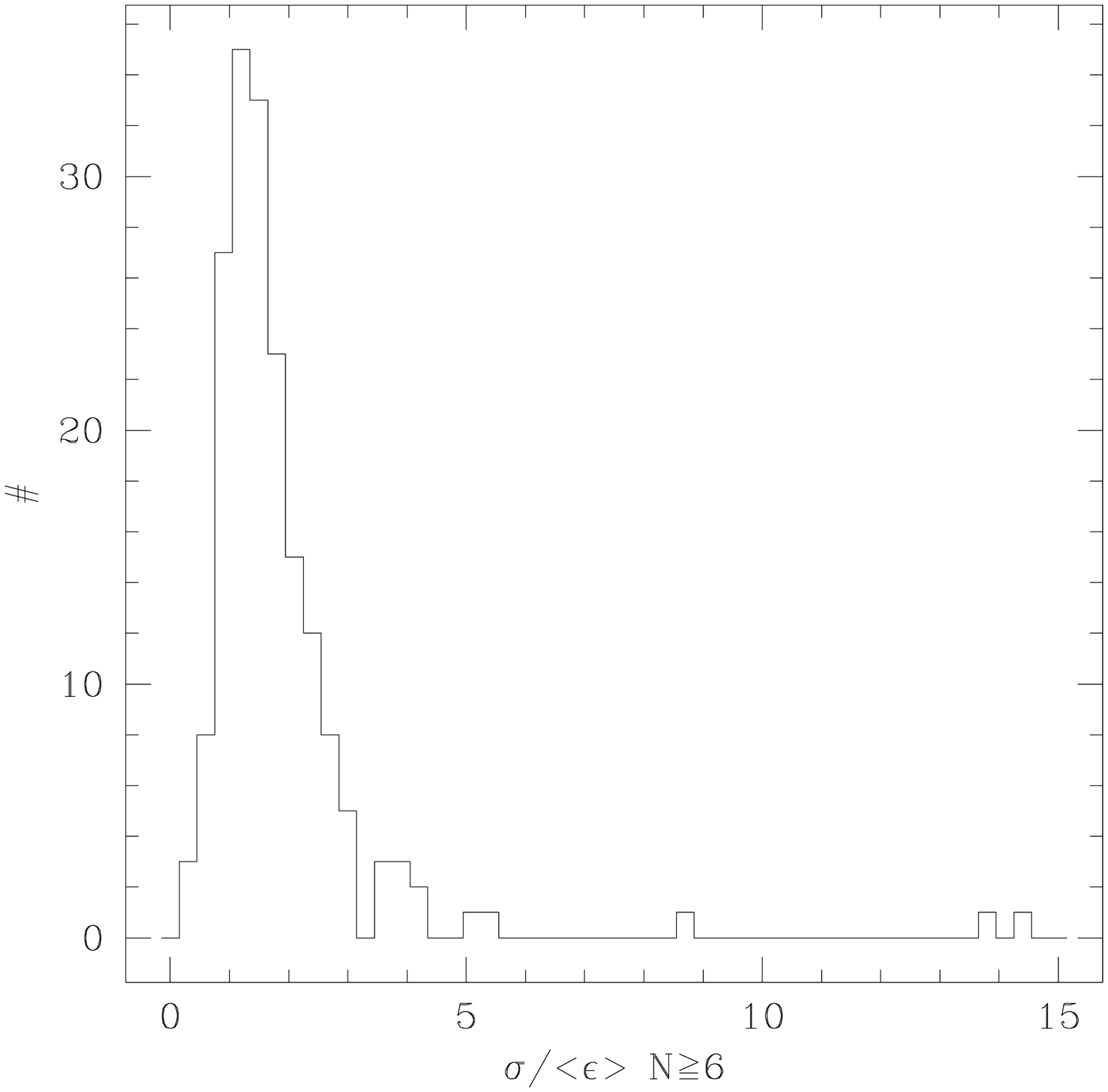}}\\
\resizebox{0.49\hsize}{!}{\includegraphics{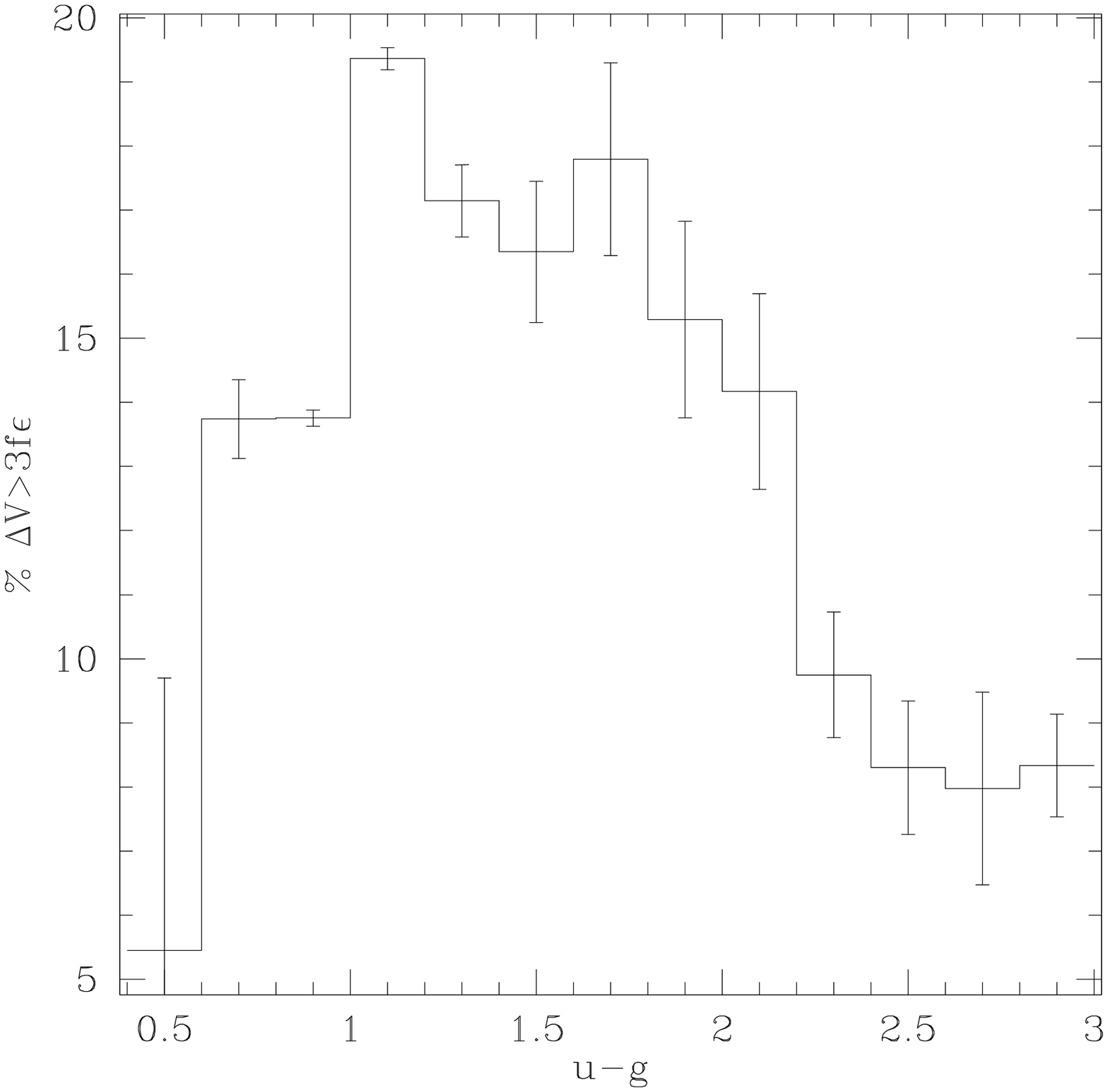}}
\resizebox{0.49\hsize}{!}{\includegraphics{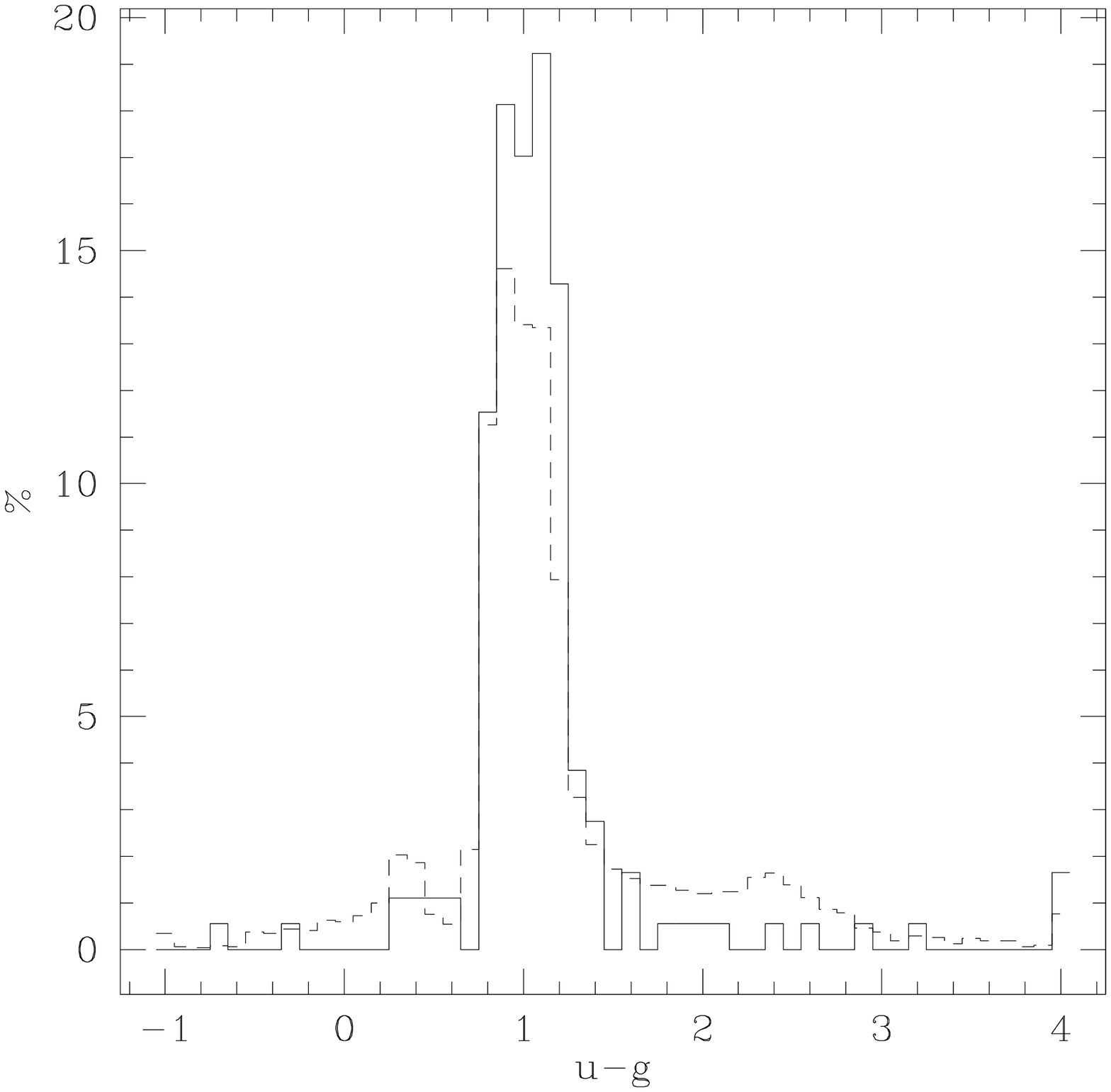}}\\
\caption[]{\label{Fig:sigRV} (Upper left panel) Ratio of the standard
deviation of the velocity of stars observed six times or more reckoned in mean
quoted velocity uncertainty versus color. (Upper right panel) Distribution
of that ratio. (Lower left Panel) Percentage of stars with 
$u-g>0.5$ and significant velocity excursions (see text). (Lower right panel)
Color distribution of the 10\,674 stars (dashed line) and those with 
significant velocity excursions (solid line). }
\end{figure}

First, we check the quoted accuracies $\epsilon$ of the radial velocities
by comparing them with the velocity dispersion $\sigma$
for multiply-observed stars.  In most of these cases,
the star has been observed only twice, but 182 stars in the
SDSS ``Southern Survey'' - a region of sky
$\rm \pm 1.25^{\circ}$ in declination along the celestial equator (J2000) between
right ascensions $\rm 20^h$ to $\rm 04^h$ - have been observed
often enough (six or more times) that a reasonable estimate can be made of the 
dispersion $\sigma$ in radial velocity.
The resulting values of $\sigma$ were compared 
with $\epsilon$ for each star, and the ratio $\sigma/\epsilon$ computed
(upper left panel of Fig.~\ref{Fig:sigRV}). This distribution is expected to 
have a tail to high values because of the presence of spectroscopic binaries 
in the sample, but if the fraction of binaries is small (see below) the median
value of $\sigma/\epsilon$ will be little affected by their presence (upper right panel of Fig.~\ref{Fig:sigRV}).  
We find the median to be 1.5, and it does not depend on the available number of observations for each star - i.e. the same median is found for stars with six or more observations, seven or more, and so on.  Thus the data suggest that the fitted velocity uncertainties $\epsilon$ may underestimate the true errors $\sigma$ by a factor of about 1.5.

Let us now turn to binary detection.  Since the
number of observations of each star is small, we search for radial
velocity variations by determining the maximum excursion of the
observed velocities, $\Delta V = V_{\rm max} - V_{\rm min}$, and asking if its
value is consistent with the quoted uncertainties. 
If $(p_1,\dots,p_n)$ are $n$ values drawn from a $N(m,\sigma)$ distribution, then
\[
\max(p_1,\dots,p_n)-\min(p_1,\dots,p_n)\sim N(0,f(n)\sigma)
\]
where $f$ depends on the number of observations, e.g. $f = \sqrt{2}$ for two
observations.  We determined $f$ for larger $n$ using 
Monte-Carlo simulations. For a given star,
the individual values of $\epsilon$ are almost always very similar, so we make the
simplifying assumption of a constant value of $\epsilon$ (just the average) for
all the observations of that star. We then conclude that a given star shows
significant velocity variations if $\Delta V > 3 f \epsilon$.  
For the 10\,647 stars with multiple observations, Table \ref{Tab:nobs} lists
the distribution of the number of observations and the corresponding values
of $f$ determined from Monte-Carlo simulations.

\begin{table}[htb]
\caption[]{\label{Tab:nobs}Number of stars ($N_{\rm obj}$) versus the
number of repeated observations ($N_{\rm obs}$).  $f$ is the
multiplier for the standard deviation
when the maximum excursion of the data is measured (see text) and grows 
with $N_{\rm obs}$.}
\begin{tabular}{lll lll lll}\hline
$N_{\rm obs}$ & $N_{\rm obj}$ & $f$ & $N_{\rm obs}$ & $N_{\rm obj}$ & $f$ & $N_{\rm obs}$ & $N_{\rm obj}$ & $f$ \\ \hline
   2   &    9108   & 1.41 & 6  &    45  & 2.68 & 10 &  24 & 3.17\\ 
   3   &     938   & 1.91 & 7  &    35  & 2.83 & 11 &   7 & 3.27\\
   4   &     287   & 2.24 & 8  &    34  & 2.96 & 12 &   9 & 3.35\\
   5   &     132   & 2.48 & 9  &    27  & 3.08 & 13 &   1 & 3.42\\
\hline
\end{tabular}
\end{table}

\begin{figure}[htb]

\resizebox{\hsize}{!}{\includegraphics{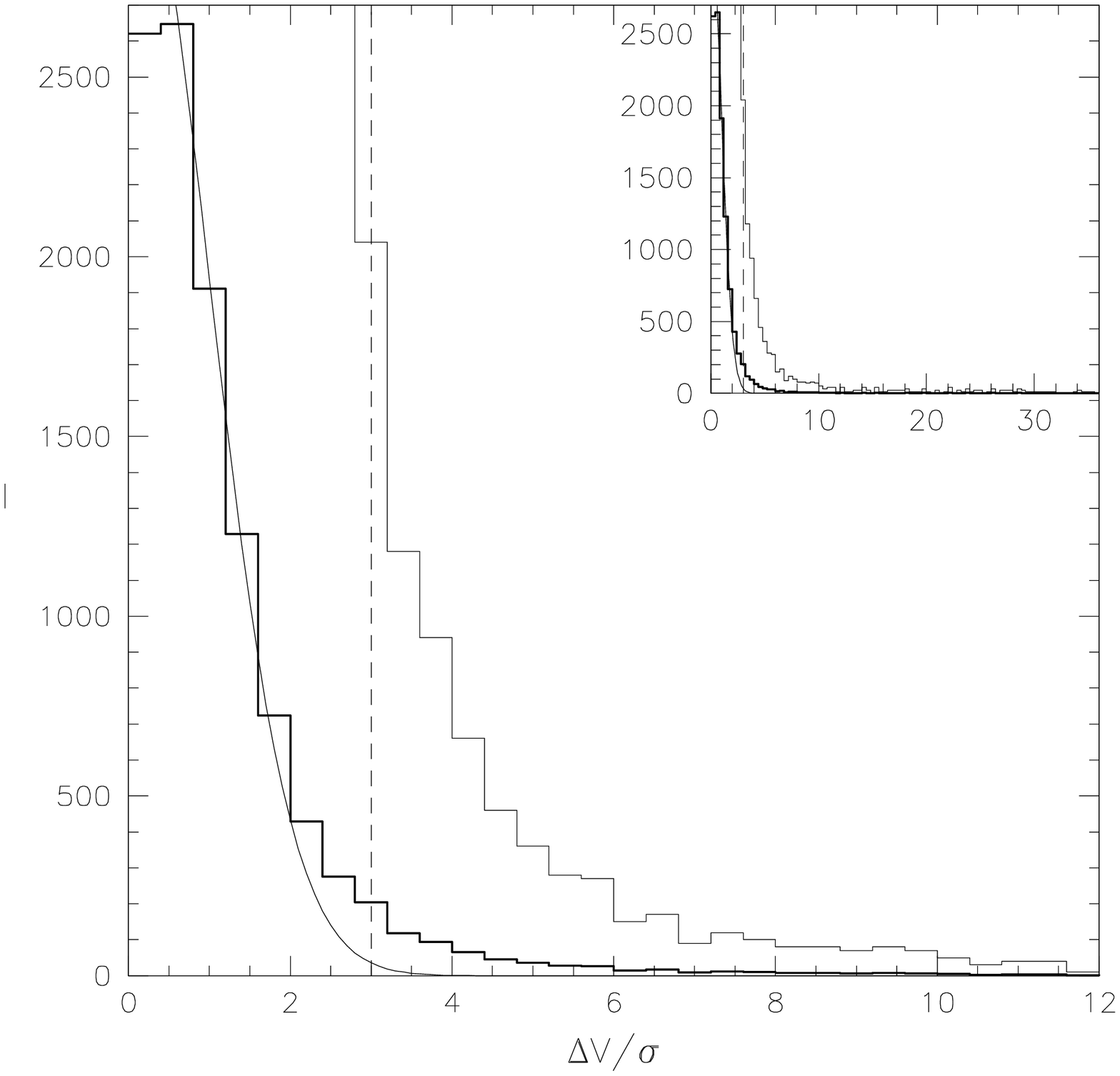}}
\caption[]{\label{Fig:Gauss} Distribution of errors for multiply-
observed stars, showing the number of stars versus the normalized
velocity excursion, $\Delta V/\sigma$ (see text).  The inset shows
the full range of this quantity. The distribution is plotted twice:
the light histogram shows the distribution with the vertical axis magnified
by 10.  The vertical dotted line shows the 3$\sigma$ criterion used 
to identify possible binary stars (see text). A Gaussian with dispersion
1 and total number of objects equal to that in the sample is compared (light
curve) with the observed distribution (histogram, heavy line).}
\end{figure}

The lower left panel of Fig.~\ref{Fig:sigRV} shows the percentage of stars for 
which $\Delta V > 3 f \epsilon$ as a function of color.  The mean percentage 
(for $0.7\le u-g\le 2.2$) is about 15\% but it does not seem to be constant 
with color. After a steep raise at $u-g\sim 1$, it decreases at a nearly constant rate up to $u-g\sim 2.7$.  It is likely that such a decrease is related to the capacity of the method to detect binaries rather than to a genuine feature of the binary distribution.  Still, this mean percentage is about 50 times 
higher than the rate expected if the radial velocities were constant and
the velocity errors Gaussian.  The 
quantity $\Delta V/1.5 f \epsilon$ was computed for the entire sample
of 10\,647 stars. The fraction of stars for which this quantity is
greater than 3 drops to 6.5\%.
Figure \ref{Fig:Gauss} shows the histogram of velocity offsets. Almost all of
the data ($>$ 90\%) are consistent with Gaussian velocity errors.

\begin{table}[htb]
\caption[]{\label{Tab:SBid}Stars for which there are enough observations to
potentially derive a spectroscopic orbit.  $N$ is the number of observations
and $\left<\epsilon \right>$ the mean of the quoted radial velocity
uncertainties (km/s).  The remaining columns list the $ugriz$ magnitudes and 
the derived spectral type Sp.}
\setlength{\tabcolsep}{1.5pt}
\begin{tabular}{ccccccccccc}\hline
Name (SDSS J.) & $N$ & $\left<\epsilon\right>$ & $u$ & $g$ & $r$ & $i$ & $z$ & Sp \\ \hline
\object{003106.81+004135.7}& 10&  3.2& 18.93& 17.79& 17.26& 17.04& 16.91& F9\\ 
\object{003546.95+001303.5}&  9&  3.1& 17.49& 16.41& 15.93& 15.77& 15.71& G2\\ 
\object{005146.88+010841.8}&  9&  2.9& 17.75& 16.56& 16.07& 15.89& 15.78& F9\\ 
\object{005406.06+003432.0}& 10&  4.3& 18.09& 17.28& 16.99& 16.88& 16.85& F2\\ 
\object{022036.00+002309.7}&  8&  3.1& 17.54& 16.44& 15.99& 15.84& 15.79& F9\\ 
\object{022216.99+000611.9}&  9&  2.1& 16.76& 15.92& 15.60& 15.50& 15.49& F2\\ 
\object{022426.98+004236.4}&  9&  2.8& 16.97& 15.90& 15.48& 15.35& 15.30& G2\\ 
\object{022502.06+001541.0}&  9&  4.4& 18.69& 17.89& 17.51& 17.39& 17.33& F2\\ 
\object{022555.65+010850.3}&  9&  2.5& 17.03& 16.20& 15.92& 15.82& 15.77& F2\\ 
\object{030225.11+010843.8}&  6&  3.2& 17.95& 16.84& 16.35& 16.15& 16.07& G2\\ 
\object{030953.46+002747.5}&  6&  4.1& 20.65& 18.80& 17.78& 17.70& 16.99& K5\\ 
\object{031404.97-011136.6}&  6&  5.7& 20.78& 19.95& 19.12& 17.80& 17.01& M3\\ 
\object{031505.31+002120.4}&  7& 14.4& 25.51& 22.27& 20.44& 19.18& 18.33& M2\\ 
\object{031540.79+002830.0}&  6&  3.2& 19.60& 18.21& 17.46& 17.15& 16.92& F9\\ 
\object{031559.14+002803.2}&  7&  9.1& 23.11& 22.54& 20.97& 18.87& 17.77& M4\\ 
\object{032937.14+011315.8}&  7&  2.7& 17.54& 16.44& 15.95& 15.74& 15.62& F5\\ 
\object{032937.49+000443.7}&  6&  5.7& 19.40& 18.52& 18.10& 17.95& 17.83& F2\\ 
\object{033209.68+005658.1}&  8&  3.1& 18.36& 17.36& 16.86& 16.67& 16.57& F5\\ 
\object{034137.67+011027.6}&  6&  3.8& 18.78& 17.63& 17.17& 16.98& 16.87& F5\\ 
\hline
\end{tabular}
\end{table}

Since we have only a few (mostly only two) observations of each star,
and these are randomly distributed in terms of the orbital
periods, we do not expect to be able to identify every binary
star even for those with velocity amplitudes several times the
SDSS velocity uncertainties. In order to estimate the detection rate.
synthetic observations with same time distribution and uncertainty as the
original observations are generated using
the compilation of orbital elements for 2\,405 spectroscopic
binary stars from
\SB9\footnote{http://sb9.astro.ulb.ac.be} \citep{Pourbaix-2004:b}.  Each one 
of the 10\,647 stars is tested against every single orbit.
The maximum velocity excursions $\Delta V$ is derived and so is the 
percentage of simulated observations
with $\Delta V >3 f \sigma$, where $\sigma$ is defined to be
1.5 times the mean quoted velocity error $\epsilon$.
This simulation shows that only 39\% of these synthetic binaries would be
identified by the existing SDSS observations.
The percentage reaches 57\% for stars observed six times or more.  Combining this result with the 6.5\% of $3\sigma$ binaries within the 
sample of stars observed more than once by SDSS which show detectable velocity 
variations, one ends up with 16.7\% of stars
in the halo which are spectroscopic binaries, very consistent with the value
of $18\%\pm 4\%$ found by \citet{Carney-2003:a}.  Owing to the effect of
the rotation on the precision of the radial velocities and, therefore, on the 
significance of the largest radial velocity difference, our inferred percentage
of binaries is a lower bound.

\section{Spectroscopic orbits}\label{Sect:orbits}

\begin{table*}[htb]
\caption[]{\label{Tab:orbits}Orbital elements and their uncertainty.
$V_0$ and $K$ are in $km/s$, $\omega$ in degrees, $P$ in days, 
epoch $T_0$ is in days + 24400000, $a_1\sin i$ in $10^6$ km and $f(M)$ in solar masses.}
\begin{tabular}{l llllll ll l}\hline
Name & $V_0$ & $e$ & $\omega$ & $K$ & $P$ & $T_0$ & $a_1\sin i$ & $f(M)$ & $\chi^2$ \\
     & $\sigma_{V_0}$ & $\sigma_{e}$ & $\sigma_{\omega}$ & $\sigma_{K}$ & $\sigma_{P}$ & $\sigma_{T_0}$ & $\sigma_{a_1\sin i}$ & $\sigma_{f(M)}$ & $F2$ \\ \hline

003106.81+004135.7& 66.80& 0.222& 271& 28.4& 14.18& 51793.7& 5.4& 0.0312& 2.26\\
&0.77& 0.057& 17& 0.7& 0.98& 0.55& 0.4& 0.0082& -0.060\\
005146.88+010841.8& -11.60& 0.343& 71& 18.4& 5.98& 51814.2& 1.4& 0.0032& 1.73\\
&0.62& 0.045& 24& 1.9& 0.58& 0.24& 0.2& 0.0015& -0.34\\
005406.06+003432.0& -61.6& 0 (fixed)& 345& 15.58& 15.9& 51806.3& 3.40& 0.0062& 14.63\\
&1.2& -- & -- & 0.62& 1.1& 0.54& 0.27& 0.0017& 2.53\\
022036.00+002309.7& 11.4& 0.234& 275& 18.6& 8.85& 51820.5& 2.2& 0.0054& 0.84\\
&1.1& 0.071& 27& 2.4& 0.35& 0.54& 0.3& 0.0023& -0.42\\
022502.06+001541.0& -56.2& 0.179& 222& 23.48& 3.57& 51819.3& 1.136& 0.00458& 2.64\\
&0.6& 0.046& 12& 0.8& 0.11& 0.13& 0.054& 0.00071& 0.12\\
030225.11+010843.8& 24.99& 0.448& 276.0& 35.8& 117.3& 51822.2& 51.7& 0.401& 6.1e-4\\
&0.87& 0.034& 5.6& 2.1& 2.3& 0.95& 3.3& 0.078& --\\
031404.97-011136.6& -52.0& 0.361& 14.8& 88& 32.8& 51873.5& 37.0& 1.90& 7.82\\
&2.5& 0.034& 8.8& 4& 1.3& 0.36& 2.3& 0.38& --\\
032937.14+011315.8& 26.26& 0 (fixed) & 97& 18.7& 21.77& 51869& 5.60& 0.0148& 0.19\\
&0.78& -- & -- & 2.1& 0.54& 2.3& 0.64& 0.0051& -0.43\\
\hline
\end{tabular}
\end{table*}

\begin{figure*}[htb]
\resizebox{0.23\hsize}{!}{\includegraphics{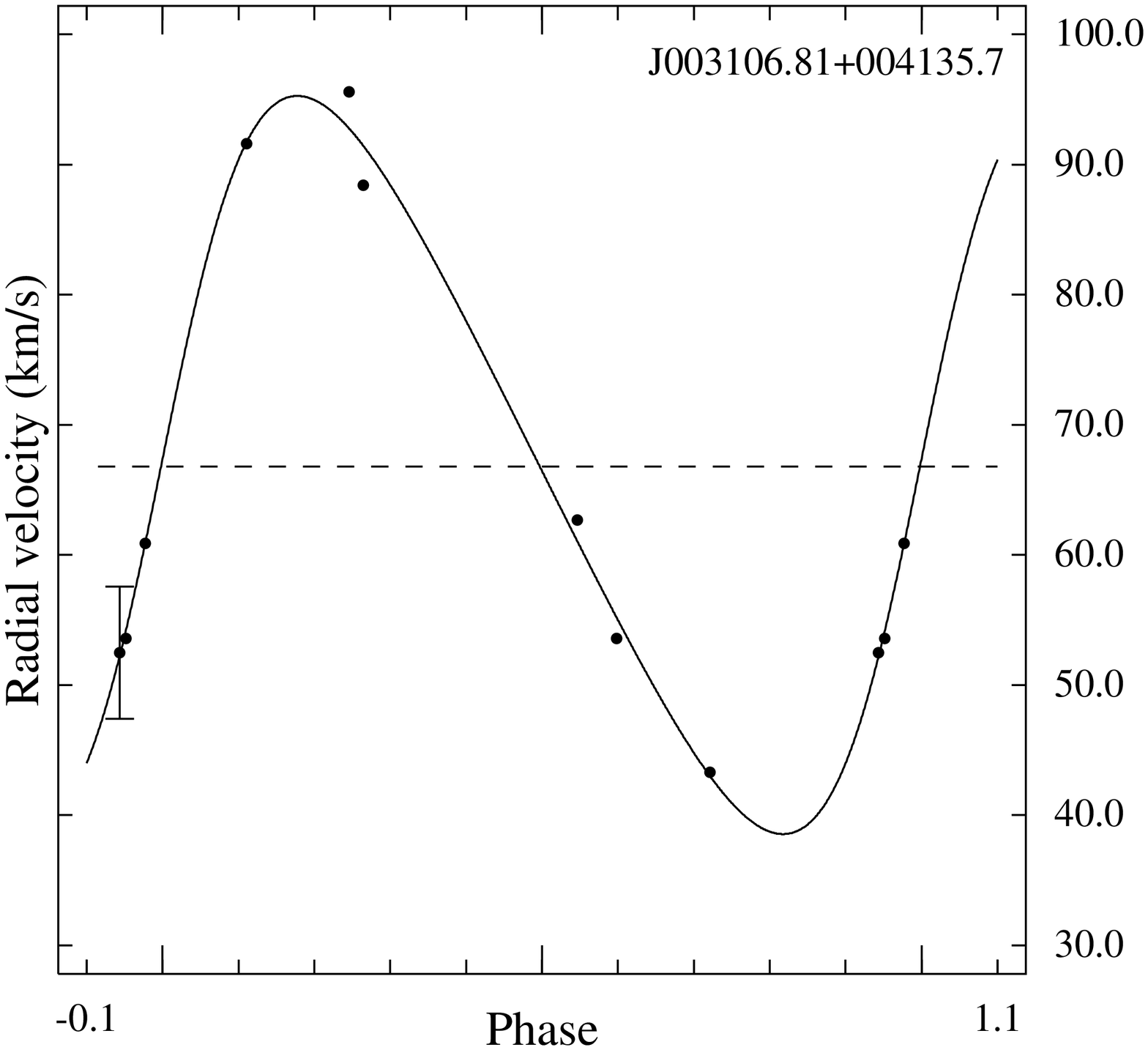}}
\resizebox{0.23\hsize}{!}{\includegraphics{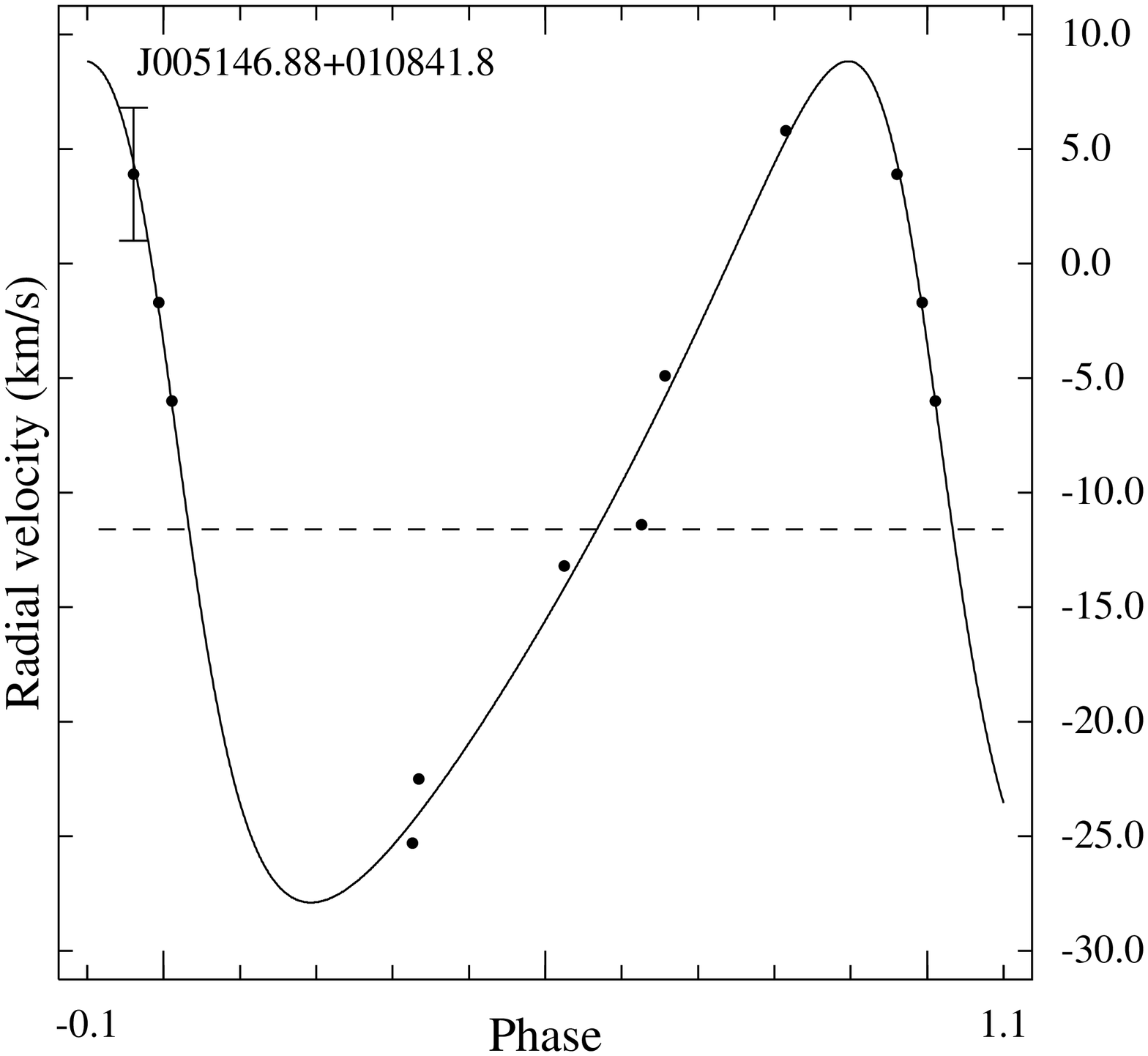}}
\resizebox{0.23\hsize}{!}{\includegraphics{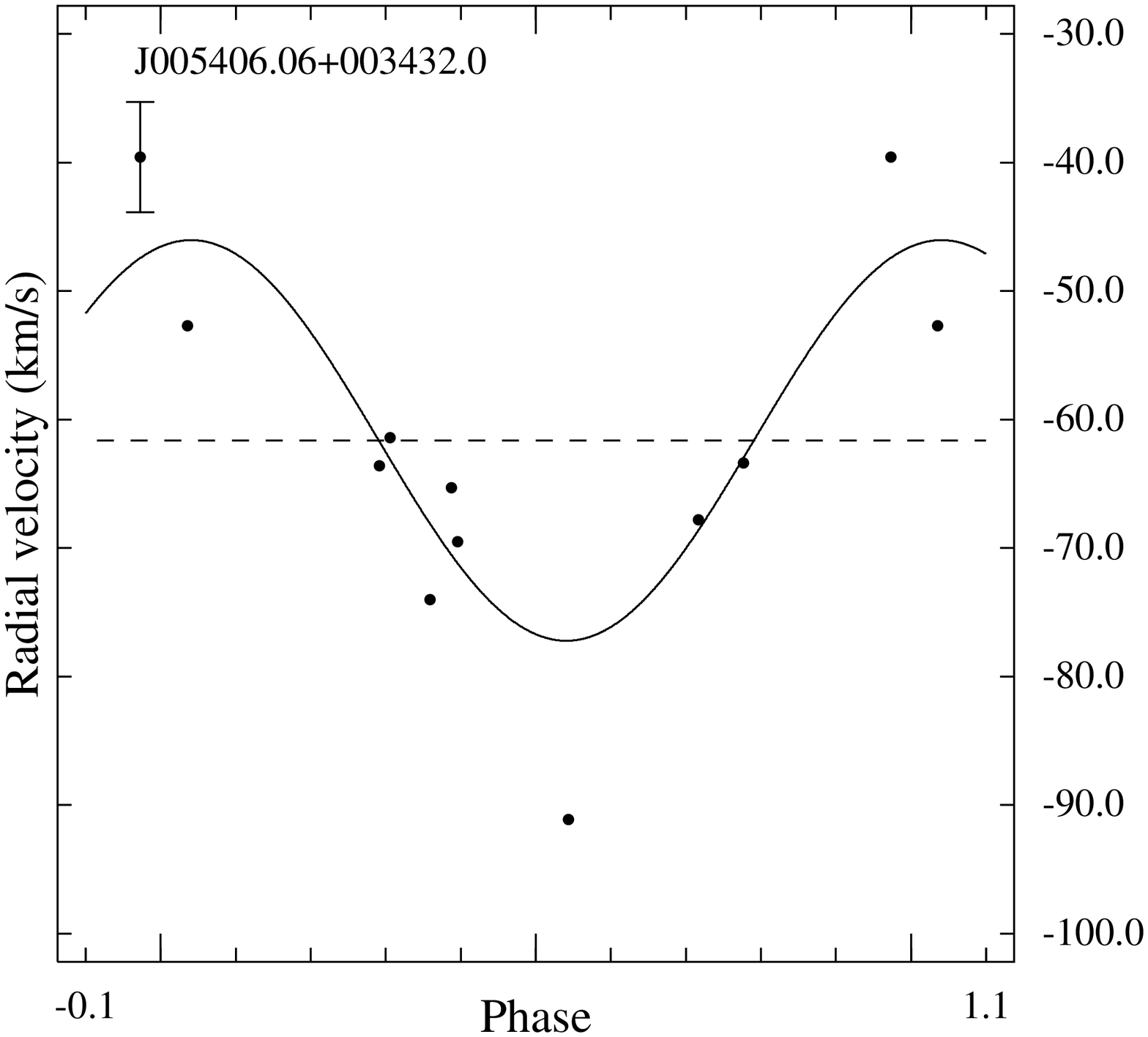}}
\resizebox{0.23\hsize}{!}{\includegraphics{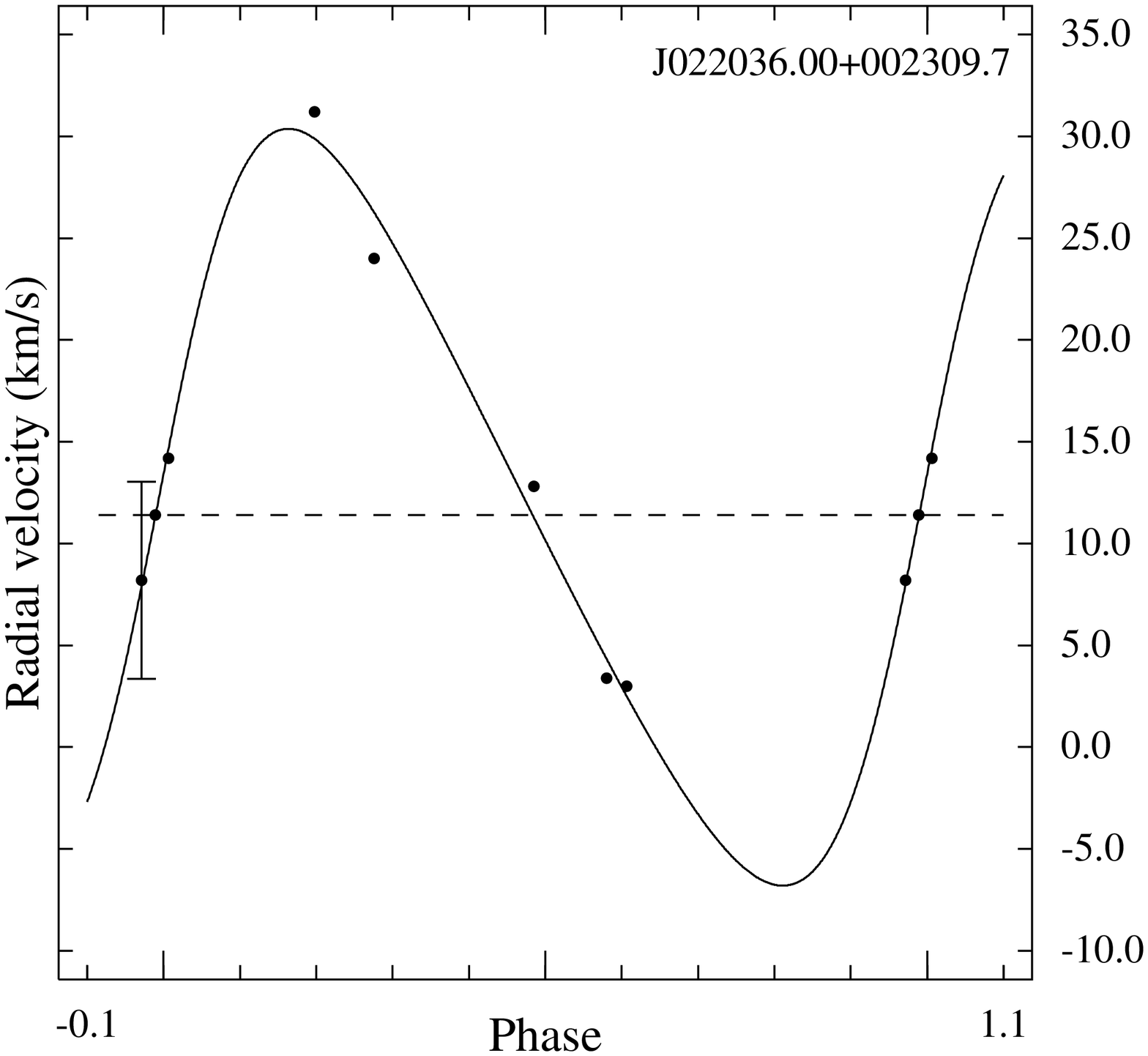}} \\
\resizebox{0.23\hsize}{!}{\includegraphics{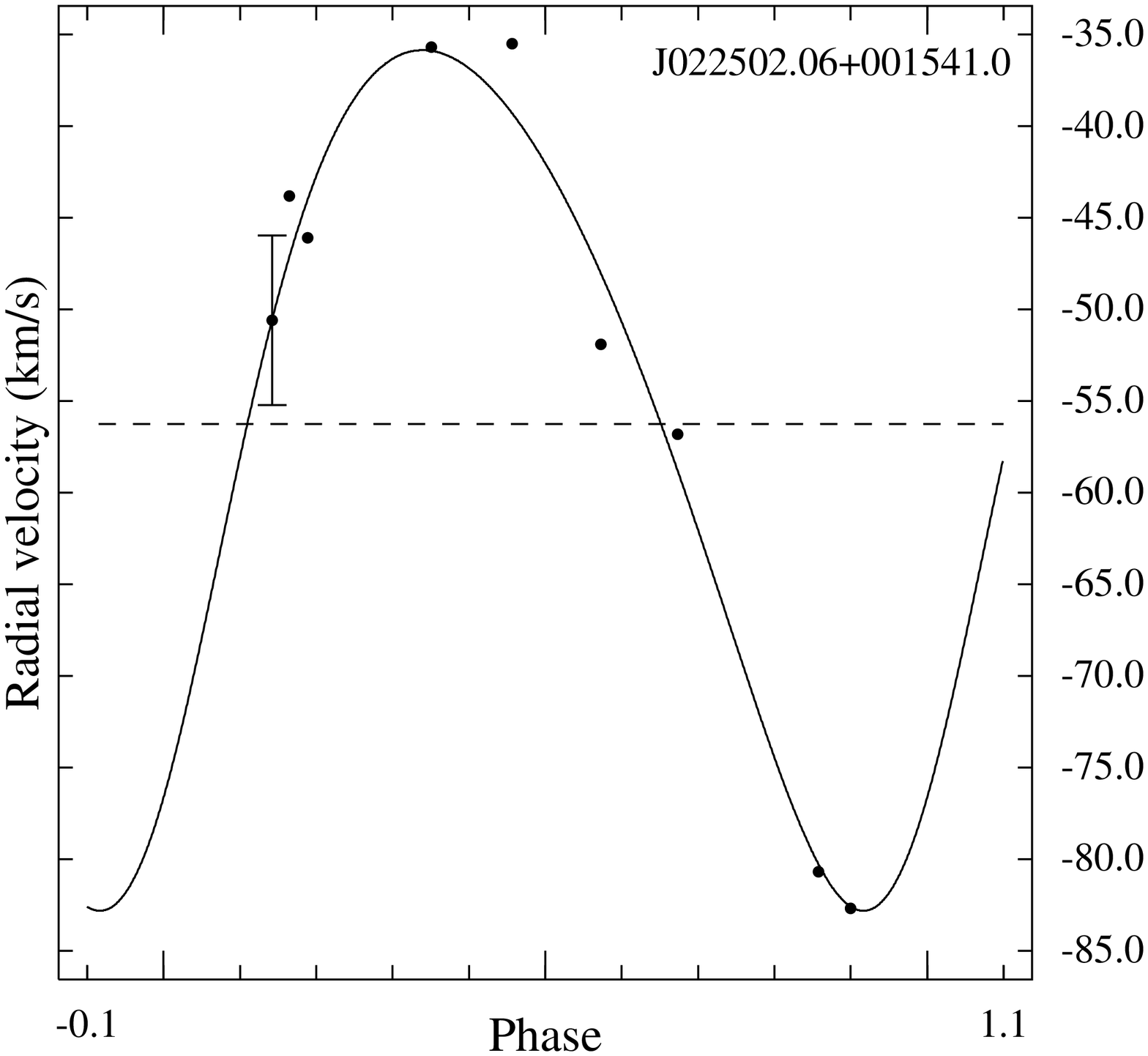}}
\resizebox{0.23\hsize}{!}{\includegraphics{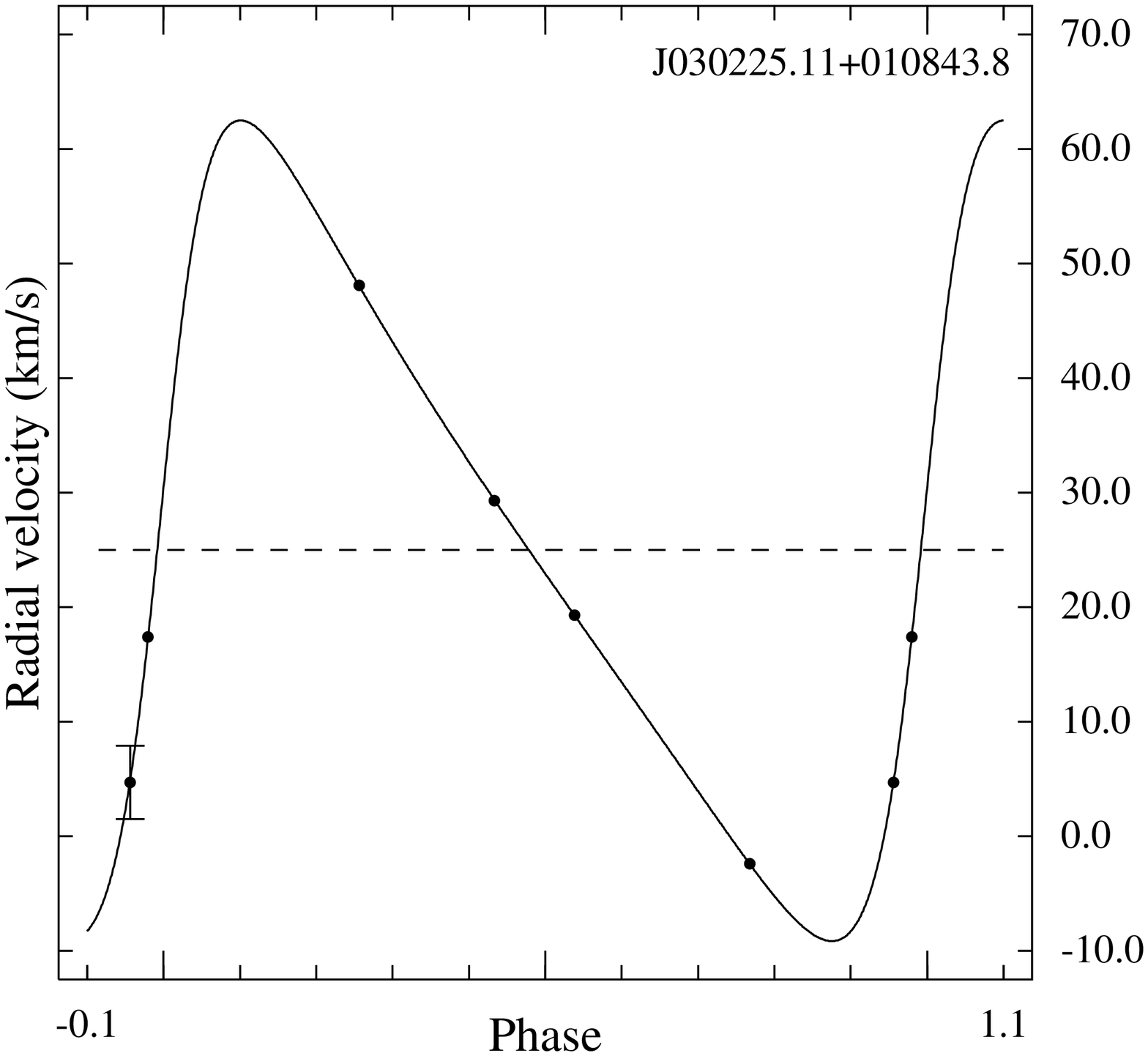}}
\resizebox{0.23\hsize}{!}{\includegraphics{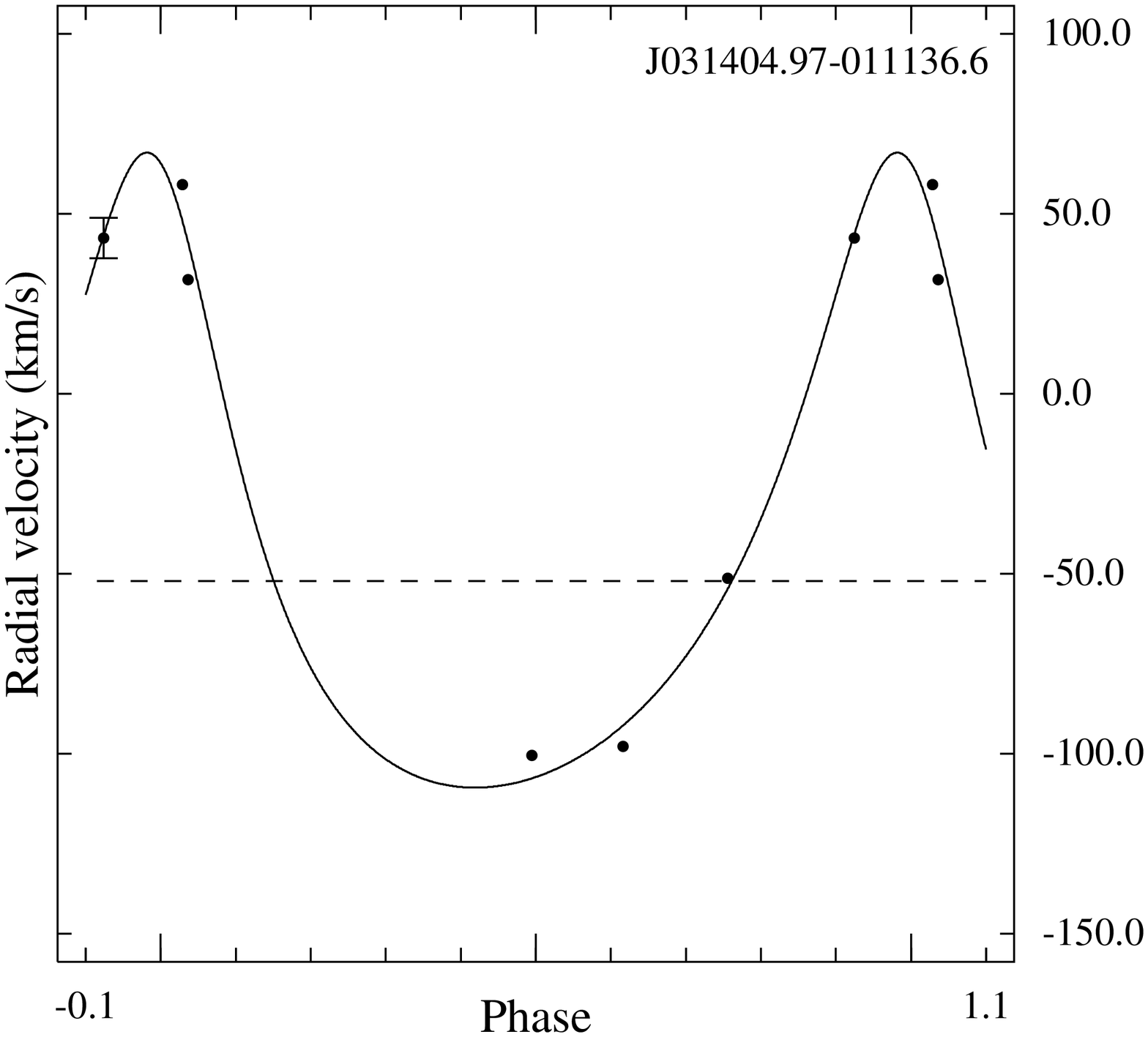}}
\resizebox{0.23\hsize}{!}{\includegraphics{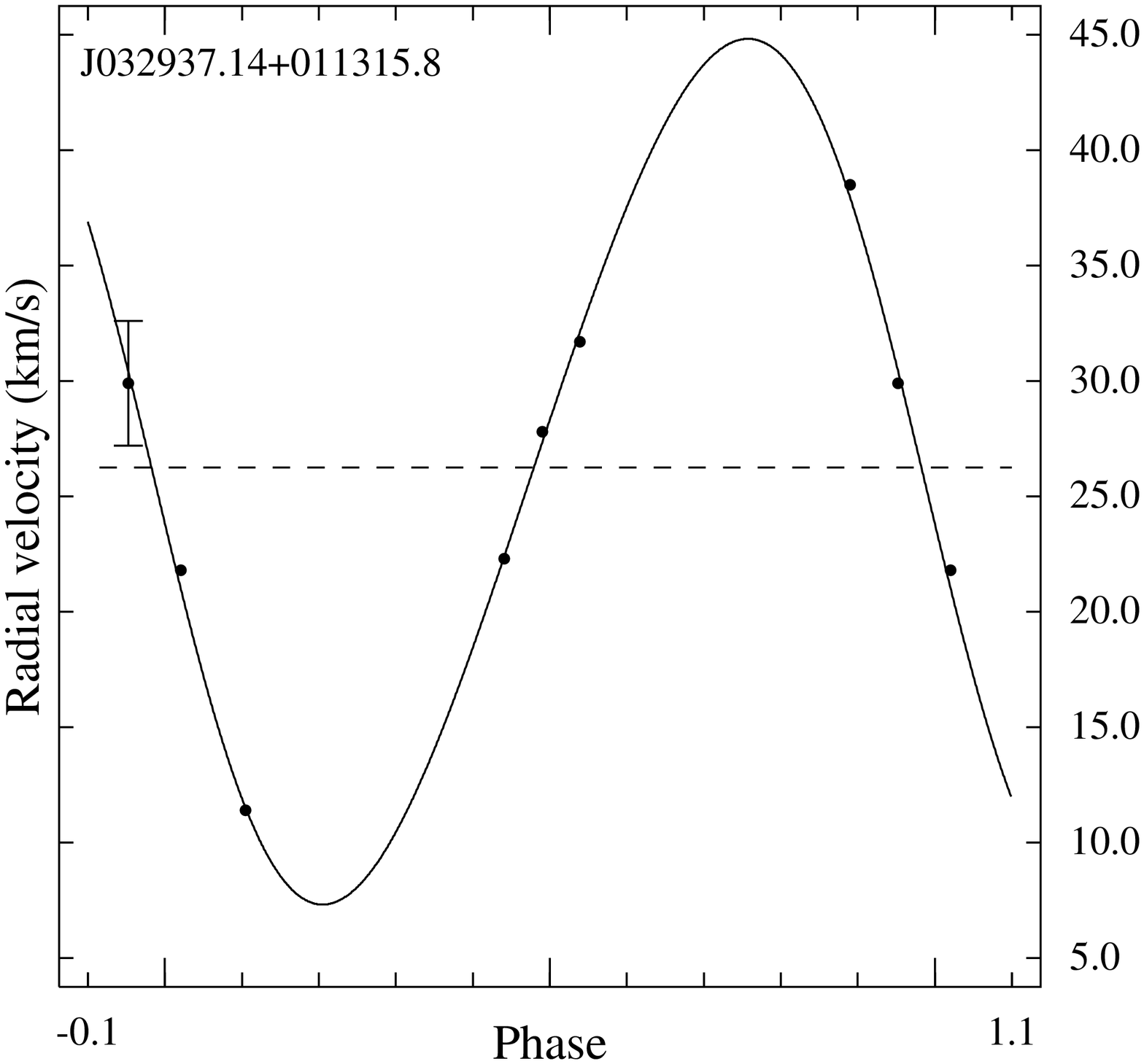}} \\
\caption[]{\label{Fig:SBplots}Plots of the orbits given in Table 
\ref{Tab:orbits}.  The dashed line shows the systemic velocity.}
\end{figure*}

Nineteen multiply-observed stars have enough observations
and a large enough velocity excursion with respect to the 
errors that an orbit can in principle be found (a minimum of
six observations is required to define an orbit, but these must
be accurately measured and well distributed with phase),
as shown by the simulations described above.
Since
the maximum number of observations per object available from 
our observations is thirteen and these are not necessarily
well distributed, a robust orbit
determination is unlikely and the orbits calculated in
this section must be regarded as preliminary. The nineteen objects
are listed in Table \ref{Tab:SBid} along with their SDSS $ugriz$
magnitudes and the spectral type assigned by the velocity fitting
program.  Upon examination of the data,
four of these were discarded: SDSS J030953.46+002747.5, J031505.31+002120.4, J031559.14+002803.2, and J032937.49+000443.7. For example, the orbit
fit to SDSS J030953.46+002747.5 is very eccentric ($e = 0.978$),
which would make it an outlier in the $e-\log P$ diagram 
\citep{Pourbaix-2004:a}, and discarding the most discrepant observation 
leaves only five points, too few for an orbital fit.
The situation for the three other stars is essentially the same.
These four stars are among the faintest of the sample
Even when the most discrepant velocity
is removed, the remaining data still exhibit a significant velocity excursion,
supporting their identification as binary stars.
For seven other stars, either the amplitude or the period is poorly 
constrained, and these preliminary solutions are not included. 

The orbits for the remaining eight stars are reasonably robust and
are given in Fig.~\ref{Fig:SBplots} and Tab. \ref{Tab:orbits}.
This table lists the name of the object, its systemic velocity $V_0$,
the eccentricity $e$, the argument of the periastron $\omega$,
the projected radial velocity amplitude $K$, 
the period $P$, and one epoch of periastron passage $T_0$.  Also
listed are the projected semi-major axis of the absolute orbit of the
primary $a_1\sin i$, the mass function $f(M)$ \citep{PrDoSt},
the value of $\chi^2$ and the goodness of fit F2 \citep{FuAs} 
\[
F2=\sqrt{9\nu\over 2}(\sqrt[3]{\chi^2\over \nu}+{2\over9\nu}-1)
\]
where $\nu$ is the number of degrees of freedom, i.e. the number
of observations minus 6.  $F2$ follows a $N(0,1)$-distribution, 
and is $<3$ for all the stars (though it is undefined when the number of 
observations equals the number of orbital parameters).
Based on the algorithm by \citet{Eyer-1999:a}, the Nyquist frequency 
is one per day, and none of the derived periods is below that limit.  The
eccentricities of the eight orbits are consistent with those of similar
systems (spectral type, orbital period) found in \SB9.
 
\begin{figure}[htb]
\resizebox{0.98\hsize}{!}{\includegraphics[angle=270]{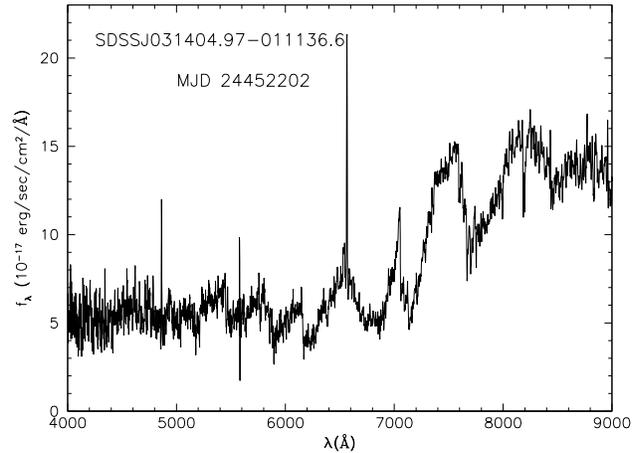}}
\caption[]{\label{Fig:spect} Spectrum of \object{SDSS J031404.97-011136.6}.
This star is an example of a white dwarf-M dwarf pair, as shown by the
blue continuum (well in excess of that expected from a dM3 star) and
the strong H$\alpha$ emission, which may be due to irradiation 
of the dM secondary by the white dwarf.}
\end{figure}

SDSS J031404.97-011136.6 has the largest signal to noise ratio of the
sample and also exhibits the largest mass function $f(M)$.  
The colors, $r-i=1.32$ and $i-z=0.78$, correspond to an M3 star 
\citep{Hawley-2002:a} and the fits to all 7 spectra also yield M3.
Examination of the spectra (one of which is shown in Fig.~\ref{Fig:spect})
shows that this is a dM/WD pair, consistent with the white
dwarf being the more massive of the pair. However, the mass 
function, $1.9\pm 0.4$ solar masses is about 
$3\sigma$ different from typical values found for these objects.

\begin{figure}[htb]
\resizebox{0.98\hsize}{!}{\includegraphics[angle=270]{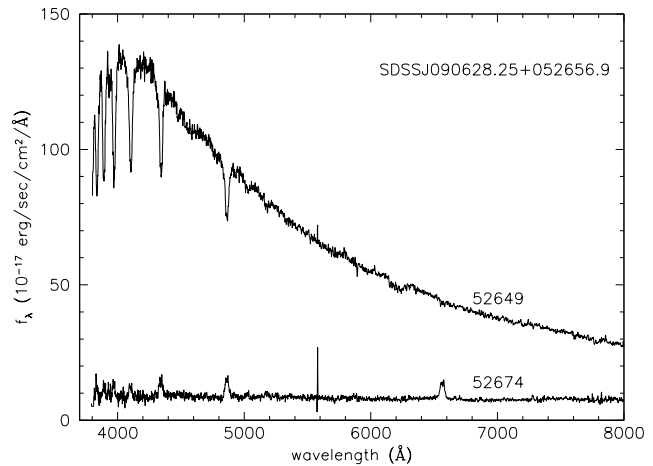}}
\caption[]{\label{Fig:spectCV} Spectra of the CV SDSS J090628.25+052656.9
labelled by MJD (-24\,400\,000) of observation. The system is observed
in both its low and high states. In the high state, rotationally
broadened Balmer absorption lines are seen, and $H\alpha$ is 
filled in by emission.}
\end{figure}

In all, we found 675 stars with velocity excursions large enough
that they are likely to be spectroscopic binaries. This list
is given in Table 4 (available only in electronic form). It includes
13 M dwarf-white dwarf (dM/WD) pairs and 7 cataclysmic variables
(CVs),
easily identified by eye examination of the spectra.
These latter are described by \citet{Szkody-2002:a,Szkody-2003:a,
Szkody-2003:b,Szkody-2004:a,Szkody-2005:a}, but one observation
of note is not included in those papers.  The second observed
spectrum of \object{SDSS J090628.25+052656.9}, on MJD 24\,452\,674, is
presented by \citet{Szkody-2005:a} and shows both red and blue
stellar components and hydrogen Balmer line emission. A spectrum
obtained a month earlier, however (MJD 24\,452\,649) catches the star
in its high state.
These spectra are shown in 
Fig.~\ref{Fig:spectCV} and are available at the SDSS web site (they can
be located using the information in Table 4). \cite{Szkody-2005:a}
note that this star is a likely dwarf nova, and the outburst spectrum
shown in Fig.~\ref{Fig:spectCV} lends support to this classification.
Further, the four observations of this object (Fig.~\ref{Fig:spectCV}
and \cite{Szkody-2005:a}) find it in outburst twice - thus the outbursts
likely repeat on a fairly short timescale.

Finally, none
of the 35 carbon stars with 
multiple observations shows significant radial velocity deviations.

\section{Conclusions}

We examine some 10\,000 stars for which multiple SDSS spectra 
have been obtained. The dispersion of the measured velocities 
is found to be about 1.5 times the quoted uncertainty of the 
radial velocity fits for stars of all observed colors (spectral
types) and magnitudes. A group of objects with large velocity 
excursions is identified and the percentage of such stars (6\%)
shown to be consistent with the expected fraction of binary stars.

We identify 675 possible new binary stars. Most of these, like most
of the observed stars, are F subdwarfs, but the list includes
13 dM/WD pairs and 7 CVs. One of these, \object{SDSS J090628.25+052656.9},
is observed in both its low and high states.
The identification of these 675 stars as binaries is very preliminary, being
based on a very small number of observations. The number of false positive
identifications is estimated to be about 40 from simulations. However, since
these stars will not be further observed by SDSS, they are presented here as
candidates for possible future study.

Eight of the stars have enough observations (6-13) and show
large enough velocity excursions with respect to the 
uncertainties that the fitted orbit is reasonably robust.
These orbits and the corresponding radial velocities are available on \SB9.

\begin{acknowledgements}
We thank the referee for many helpful comments which significantly improved
the paper.
Partial support for the computer systems required to process and store 
the data was provided by NASA via grant NAG5-6734 and by Princeton 
University. DP thanks the American Astronomical Society for the award
of a Chr\'etien International Research Grant.  We also thank Princeton
University for generous support.  This research made use of the IDL
Astronomy User's Library at Goddard.

Funding for the creation and distribution of the SDSS Archive has been 
provided by the Alfred P. Sloan Foundation, the Participating 
Institutions, the National Aeronautics and Space Administration, the
National Science Foundation, the U.S. Department of Energy, the
Japanese Monbukagakusho, and the Max Planck Society. The SDSS Web site
is http://www.sdss.org/. The SDSS is managed by the Astrophysical
Research Consortium (ARC) for the Participating Institutions. The
Participating Institutions are The University of Chicago, Fermilab, 
the Institute for Advanced Study, the Japan Participation Group,
The Johns Hopkins University, the Korean Scientist Group, Los Alamos
National Laboratory, the Max-Planck-Institute for Astronomy (MPIA), 
the Max-Planck-Institute for Astrophysics (MPA), New Mexico State 
University, University of Pittsburgh, University of Portsmouth,
Princeton University, the United States Naval Observatory, and 
the University of Washington.
\end{acknowledgements}

\bibliographystyle{aa} 
\bibliography{articles,books}

\end{document}